%% file: main.tex
\documentclass[journal]{new-aiaa}
\usepackage{textcomp}

\usepackage{graphicx}
\usepackage{amsmath}
\usepackage[version=4]{mhchem}
\usepackage{siunitx}
\usepackage{longtable,tabularx}
\usepackage{hyperref}
\usepackage{ragged2e}
\hypersetup{
    hidelinks
}

\usepackage{url}
\usepackage{subcaption}
\usepackage{enumitem}
\usepackage{amsmath}
\usepackage{bm}

\usepackage{multirow}
\usepackage{ulem}

\usepackage[title]{appendix}
\usepackage{array}   

\usepackage{lipsum} 
\usepackage{xcolor} 
\usepackage{todonotes}

\input{annotation.tex}
\usepackage{lineno} 

\usepackage{changes}
\definechangesauthor[name={Reviewer 1}, color = blue]{R1}
\definechangesauthor[name={Editor}, color = red]{R2}
\definechangesauthor[name={Liu}, color = blue]{ly}

\usepackage{siunitx}
\usepackage{longtable,tabularx}
\title{Evaluation of Statistical Consistency in Synthetic Turbulence under Wavenumber Bounds}

\author{Hongyuan Lin \footnote{Postdoctoral Research Associate, LNM, Institute of Mechanics; linhongyuan@imech.ac.cn. }}

\author{Yi Liu \footnote{ Postdoctoral Research Associate, LNM, Institute of Mechanics; liuyi@imech.ac.cn.}}

\author{Shizhao Wang \footnote{Professor, LNM, Institute of Mechanics; wangsz@lnm.imech.ac.cn (Corresponding author).}}

\affil{State Key Laboratory of Nonlinear Mechanics, Chinese Academy of Sciences, 100190 Beijing, \\ People’s Republic of China}

\author{Chun-Hian Lee\footnote{Professor, National Laboratory for Computational Fluid Dynamics, School of Aeronautic Science and Engineering; lichx@buaa.edu.cn.}}
\affil{Beihang University, Beijing, People's Republic of China, 100191}

\normalsize

\begin{document}

\maketitle

    \begin{abstract}
    The random Fourier method (RFM) is widely employed for synthetic turbulence due to its mathematical clarity and simplicity. However, deviations remain between prescribed inputs and synthetic results, and the origin of these errors has not been fully evaluated. 
    This study aims to systematically evaluate the effects of spectral coefficient calibration, grid constraints, and wavenumber bounds on the accuracy of RFM-generated turbulence.
    The results show that the recalibration of spectral coefficients is essential to ensure consistency in turbulent kinetic energy. The upper wavenumber bound, determined by grid resolution, controls the overall turbulent kinetic energy level, whereas the lower bound, set by computational domain size, governs the fidelity of the energy spectrum in the low-wavenumber range. Moreover, extending wavenumber bounds that exceed the grid-constrained bounds may improve turbulent kinetic energy accuracy but simultaneously amplifies spectral deviations.
    These findings clarify the distinct roles of coefficient calibration and wavenumber bounds in ensuring the statistical consistency of synthetic turbulence, providing practical guidance for selecting parameters in computational fluid dynamics and computational aeroacoustics applications.
    
    \end{abstract}

\section*{Nomenclature}

{
\renewcommand\arraystretch{1.0}
\noindent\begin{longtable*}{@{}l @{\quad=\quad} l@{}}
$\alpha$           & Spectral coefficient in von Kármán--Pao spectrum \\
$E(\kappa)$        & Energy spectrum at wavenumber $\kappa$, m$^{3}$/s$^{2}$ \\
$\mathit{Err}$     & Relative error \\
$\bm{\kappa}$      & Wavevector \\
$\kappa$           & Wavenumber magnitude, m$^{-1}$ \\
$L$                & Integral length scale or domain size, m \\
$L_e$              & Maximum eddy length, m \\
$M$                & Number of Fourier modes \\
$N$                & Number of quadrature or grid points \\
$Re_{L}$           & Local Reynolds number \\
$u'$               & RMS velocity fluctuation (single component), m/s \\
$u_m$              & Amplitude of $m$-th Fourier mode, m/s \\
$u_{\mathrm{rms}}, v_{\mathrm{rms}}, w_{\mathrm{rms}}$ & RMS velocity in $x$-, $y$-, $z$-dir., m/s \\
$\bm{u}'$          & Fluctuating velocity vector, m/s \\
$\bm{x}$           & Position vector, m \\
$\Delta x$         & Grid spacing, m \\
$\Delta \kappa_m$  & Wavenumber interval of $m$-th mode, m$^{-1}$ \\
$\varepsilon$      & Dissipation rate, m$^{2}$/s$^{3}$ \\
$\nu$              & Kinematic viscosity, m$^{2}$/s \\
$\psi_m$           & Random phase angle of $m$-th mode \\
$\bm{\sigma}_m$    & Divergence-free unit vector of $m$-th mode \\
$\mathit{TKE}$     & Turbulent kinetic energy, m$^{2}$/s$^{2}$ \\
$U(a,b,z)$         & Confluent hypergeometric function (second kind) \\
$\omega$           & Turbulence frequency scale, s$^{-1}$ \\
$z$                & Auxiliary variable for coefficient calibration \\
$\Gamma(x)$        & Euler gamma function \\

\multicolumn{2}{@{}l}{Subscripts} \\
$i, j, m$          & Indices of discrete point or Fourier mode \\
mean               & Mean value \\
pre                & Prescribed target value \\
syn                & Synthetic turbulence \\
rms                & Root-mean-square value \\
$\min$             & Lower wavenumber bound \\
$\max$             & Upper wavenumber bound \\
$e$                & Energy-containing wavenumber \\
$\eta$             & Kolmogorov wavenumber \\

\end{longtable*}
}

\setcounter{table}{0}

\section{Introduction}
Synthetic turbulence is of central importance in computational fluid dynamics and computational aeroacoustics \cite{Wu2017,Dieste2012}. 
Its applications include broadband noise studies \cite{Dieste2012,Kissner2019,Ewert2011}, turbulent inflow generation for high-fidelity simulations \cite{Wu2017,Treleaven2020,Dhamankar2017}, and particle dispersion researches \cite{Kraichnan1970,Zhou2019,Robinson2012}.
If not generated with sufficient accuracy, synthetic turbulence can introduce artificial noise artifacts or distort the physical mechanisms under investigation \cite{Treleaven2020}. 
Therefore, generating synthetic turbulence with prescribed statistical properties is a critical challenge.

Over the past decades, numerous synthetic turbulence generation techniques have been proposed, including the Random Fourier Method (RFM) \cite{Dhamankar2017,Kraichnan1970,Yu2014}, 
the Synthetic Eddy Method \cite{Jarrin2006,kim2015}, and the Digital Filtering Method \cite{Dieste2012,Treleaven2020,shen2019}. 
Among these, the RFM, first introduced by Kraichnan \cite{Kraichnan1970}, has achieved broad adoption owing to its conceptual simplicity and convenience \cite{Wu2017,Saad2017}. It is worth noting that the RFM exhibits two major deficiencies in terms of statistical consistency: (i) errors at the energy level, manifested as discrepancies in the total turbulent kinetic energy ($TKE$) \cite{Guglielmi2025}; and (ii) errors at the spectral level, manifested as distortions in the turbulent energy spectrum ($E(\kappa)$) within the discretized wavenumber range \cite{Saad2017}.

Regarding $\mathit{TKE}$ consistency, both the spectral coefficient of the turbulence energy spectrum and the admissible wavenumber range are of critical importance. 
By incorporating Pao’s spectral correction \cite{Pao1965,Pao1968} into the von Kármán spectrum \cite{Karman1948}, Karweit et al.~\cite{Karweit1991} established the widely adopted von Kármán–Pao spectrum, which provides explicit relations between spectrum and wavenumber amplitudes.
For this spectrum, Béchara et al. \cite{Bechara1994} further expressed the coefficient through $\mathit{TKE}$ and dissipation constraints but did not provide explicit values. 
Bailly and Juv\'{e} \cite{Bailly1999} assumed an infinite Reynolds number, neglecting the exponential term, and proposed specific coefficients. 
More recently, Guglielmi et al.~\cite{Guglielmi2025} argued that such an assumption is not always locally valid and introduced a tuning equation for calculating coefficients at each grid point, thereby achieving closer agreement between synthetic and target $\mathit{TKE}$. Nevertheless, agreement in $\mathit{TKE}$ alone does not guarantee correct spectral representation, particularly under grid constraints.

The consistency of $E{(\kappa)}$ is primarily governed by the wavenumber range. Various criteria for selecting its lower and upper bounds, denoted by $\kappa_\mathrm{min}$ and $\kappa_\mathrm{max}$, are summarized in Table~\ref{tab:wavenumber}.
Karweit et al. \cite{Karweit1991} restricted the range based on experimental data; 
Béchara et al. \cite{Bechara1994} employed the integral scale and the Kolmogorov wavenumber; 
Bailly and Juv\'{e} \cite{Bailly1999} linked the range to computational domain size and grid resolution; 
Casalino and Barbarino \cite{Casalino2003} proposed an empirical coefficient method.
Most of these studies provide empirical ranges for synthetic turbulence but lack analysis of their validity and the underlying error origins, thereby limiting the accuracy of their application.

Saad et al. \cite{Saad2017,Saad2016} further related the wavenumber range to sampling theory, explicitly connecting the range to grid constraints. However, their analysis did not incorporate spectral coefficient calibration, resulting in substantial discrepancies in $\mathit{TKE}$. Furthermore, they reported two issues that will be demonstrated in the present work to arise from the choice of wavenumber range: 
(i) a persistent energy deficit near the integral scale that cannot be eliminated by increasing the number of modes or adopting logarithmic distributions; and
(ii) a non-monotonic dependence of spectral error on mode number, with errors increasing again beyond $10^4$ modes and remaining within 5–10\%.

\begin{table}[htbp]
\caption{Wavenumber range selection criteria for the Random Fourier Method in literature.}
\label{tab:wavenumber}
\centering
\begin{tabular}{cccc >{\RaggedRight\arraybackslash}p{5cm}}
\hline\hline
Year & Reference & $\kappa_\mathrm{min}$ & $\kappa_\mathrm{max}$ & Remarks \\
\hline
1991 & Karweit et al. \cite{Karweit1991} & $1 \,\mathrm{m}^{-1}$ & $1000 \,\mathrm{m}^{-1}$ & Suitable for the specific experiment. \\
1994 & Béchara et al. \cite{Bechara1994} & $2\pi/L_e$ & $\kappa_\eta$ & $L_e$: maximum eddy length; $\kappa_\eta$: Kolmogorov wavenumber. \\
1999 & Bailly \& Juv\'{e} \cite{Bailly1999} & $\kappa_{e}/5$ & $2\pi/(7\Delta x)$ & $\kappa_e$: energy-containing wavenumber; $\Delta x$: grid spacing. \\
2003 & Casalino \& Barbarino \cite{Casalino2003} & $c_2 \kappa_e$ & $\min(c_3 \kappa_e, \kappa_\eta, 2\pi/(6\Delta x))$ & $c_2$, $c_3$: empirical coefficients. \\
2017 & Saad et al. \cite{Saad2017} & $2\pi/L$ & $\pi/\Delta x$ & $L$: computational domain size. \\
\hline\hline
\end{tabular}
\end{table}

In summary, while prior studies have addressed coefficient determination or proposed empirical wavenumber criteria, little attention has been paid to error origins, which may lead to unexpected discrepancies in practical applications.
In this work, we systematically evaluate the effects of spectral coefficient calibration, grid constraints, and wavenumber bounds on the statistical consistency of RFM-generated turbulence. The results reveal that $\mathit{TKE}$ errors mainly originate from high-wavenumber truncation, whereas $E(\kappa)$ errors arise from insufficient low-wavenumber coverage, and emphasize that wavenumber selection must consider grid constraints to avoid nonphysical discrepancies. These results clarify the error origins of the RFM and offer practical guidance for parameter selection to enhance statistical consistency in computational fluid dynamics and computational aeroacoustics applications.

The remainder of this paper is organized as follows. Section~\ref{sec:2-Theory} introduces the theoretical framework and error definitions.
The test cases and corresponding results are discussed in Section~\ref{sec:3-Results&Discussion}.
Finally, conclusions are drawn in Section~\ref{sec:4-Conclusion}.

\section{Random Fourier Method for Synthetic Turbulence}
\label{sec:2-Theory}

\subsection{Fundamentals of the Random Fourier Method}
The objective of the Random Fourier Method for synthetic turbulence generation is to construct a fluctuating velocity field that satisfies prescribed statistical constraints on a discrete grid, such that the target energy spectrum and the turbulent kinetic energy are reproduced with high fidelity. In practice, the method synthesizes turbulence by superposing a finite set of random Fourier modes. First, the target energy spectrum is discretized into a set of finite wavenumber bands, and a modal amplitude is assigned to each band. Next, the modal amplitudes are associated with wavevectors. Finally, all modes are superposed with randomly assigned phases to obtain the fluctuating velocity field in physical space. The fluctuating velocity component $\boldsymbol{u}'$ at an arbitrary spatial location $\bm{x}$ can be expressed as follows, based on the formulations presented by Béchara et al.~\cite{Bechara1994} and Saad et al.~\cite{Saad2017}:
\begin{equation}
\bm{u}'(\bm{x}) = 2\sum_{m=1}^{M} u_m \cos\!\big(\bm{\kappa_m} \cdot \bm{x} + \psi_m\big)\bm{\sigma_{m}},
\end{equation}
where $M$ is the total number of modes, $u_m$ is the amplitude of the $m$th mode, 
$\bm{\kappa_m}$ is the $m$th wavevector with the random distribution on the wavenumber shell $\kappa_m$, $\psi_m$ is the random phase angle, and $\bm{\sigma_m}$ is the unit direction vector. The direction of $\bm{\kappa}_m$ is described by two angles, the azimuthal angle $\varphi_m$ and the polar angle $\theta_m$, which follow the probability density functions ~\cite{Bechara1994}:
\begin{equation}
P(\varphi_m) = \frac{1}{2\pi}, \qquad 0 \leq \varphi_m < 2\pi,
\end{equation}
\begin{equation}
P(\theta_m) = \tfrac{1}{2}\sin(\theta_m), \qquad 0 \leq \theta_m \leq \pi.
\end{equation}
The vector $\bm{\sigma}_m$ lies in the plane orthogonal to $\bm{\kappa}_m$ to ensure the divergence-free condition. 
It is assumed that $\bm{\sigma}_m$ is uniformly distributed in this plane. Accordingly, the orientation angle $\alpha_m$ between $\bm{\sigma}_m$ and a reference axis in the plane is chosen with the uniform probability density
\begin{equation}
P(\alpha_m) = \frac{1}{2\pi}, \qquad 0 \leq \alpha_m < 2\pi.
\end{equation}
The phase angle $\psi_m$ is also taken as uniformly distributed,
\begin{equation}
P(\psi_m) = \frac{1}{2\pi}, \qquad 0 \leq \psi_m < 2\pi.
\end{equation}
The amplitude $u_m$ is determined from the prescribed turbulence energy spectrum $E(\kappa)$ and the wavenumber interval $\Delta \kappa_m$, namely $u_m = \sqrt{E(\kappa)\Delta\kappa_m}$. A widely used spectrum form is the von K\'arm\'an-Pao spectrum~\cite{Bailly1999}:
\begin{equation}
E(\kappa) = \alpha \frac{u'^2}{\kappa_e} \frac{\left( \kappa/\kappa_e \right)^4}{\left[ 1 + \left( \kappa/\kappa_e \right)^2 \right]^{17/6}} \exp\left[ -2 \left( \kappa/\kappa_\eta \right)^2 \right]
\label{eq:spectrum}
\end{equation}
with $\alpha$ denoting the spectral coefficient, $\kappa$ the wavenumber magnitude, $\kappa_e$ the energy-containing wavenumber, and $\kappa_\eta = \varepsilon^{1/4}\nu^{-3/4}$ the Kolmogorov wavenumber. Here, $\varepsilon$ is the dissipation rate and $\nu$ is the kinematic viscosity. The values of $\alpha$ and $\kappa_e$ will be obtained from integral constraints that connect the target energy spectrum to both $\mathit{TKE}$ and $\varepsilon$~\cite{Bailly1999}.

\subsection{Calibration of Spectral Coefficients}
\label{sec:2.2}

Bailly and Juv\'{e}~\cite{Bailly1999} determined the spectral coefficient $\alpha$ and the energy-containing wavenumber $\kappa_{e}$ by assuming an infinite Reynolds number, i.e., that the separation between the energy-containing and dissipative scales is unbounded, implying $\kappa_{e}/\kappa_{\eta}\to 0$. Under this assumption, $\alpha \approx 1.453$ and $\kappa_{e} \approx 0.747/L$, where $L$ is the integral length scale inferred from $\mathit{TKE}$ and $\varepsilon$. To account for finite Reynolds number effects, Guglielmi et al.~\cite{Guglielmi2025} introduced the local Reynolds number
\begin{equation}
Re_{L}=\frac{\mathit{TKE}^{2}}{\varepsilon \nu}
\label{eq:ReL}
\end{equation}
and the auxiliary parameter
\begin{equation}
z = 2\left(\frac{\kappa_{e}}{\kappa_{\eta}}\right)^{2}.
\end{equation}
Using the confluent hypergeometric function of the second kind,
\begin{equation}
U(a,b,z)=\frac{1}{\Gamma(a)}\int_{0}^{\infty} 
e^{-zt}\, t^{a-1}(1+t)^{\,b-a-1}\,\mathrm{d}t,
\end{equation}
where the Euler gamma function is
\begin{equation}
\Gamma(x)=\int_{0}^{\infty} t^{x-1}e^{-t}\,\mathrm{d}t,
\end{equation}
this leads to the relation
\begin{equation}
\frac{U\!\left(\tfrac{7}{2},\,\tfrac{5}{3},\,z\right)}
     {U\!\left(\tfrac{5}{2},\,\tfrac{2}{3},\,z\right)}
= \frac{2}{5}\,Re_{L}^{-1/2}.
\label{eq:z}
\end{equation}
Solving Eq.~\eqref{eq:z} numerically for $z$, the coefficients $\alpha$ and $\kappa_{e}$ are obtained as
\begin{equation}
\alpha=\frac{4}{\sqrt{\pi}\,U\!\left(\tfrac{5}{2},\,\tfrac{2}{3},\,z\right)},
\qquad
\kappa_{e}=\kappa_{\eta}\sqrt{\tfrac{z}{2}}.
\label{eq:coef}
\end{equation}

\subsection{Error Definitions}

While spectral coefficient correction theoretically improves agreement between synthetic turbulence statistics and prescribed targets, additional errors arise from grid constraints, which include both the grid resolution and the domain size. For a uniform Cartesian grid with spacing $\Delta x$ and domain size $L$, Saad et al.~\cite{Saad2017} showed that the grid-constrained wavenumber bounds are
\begin{equation}
\kappa_{\min} = \frac{2\pi}{L}, 
\qquad 
\kappa_{\max} = \frac{\pi}{\Delta x},
\label{eq:KappaRange}
\end{equation}
where $\kappa_{\min}$ is constrained by the computational domain size, and $\kappa_{\max}$ is determined by the grid resolution via the Nyquist wavenumber.

The discrepancies between the synthetic turbulence field and the prescribed targets manifest primarily in two aspects: the overall turbulent kinetic energy $\mathit{TKE}$ and the distribution of the energy spectrum $E{(\kappa)}$. To quantify these deviations, the following error measures are used.

First, the relative error in $\mathit{TKE}$ is defined as
\begin{equation}
\mathit{Err}_{\mathit{TKE}} = 
\frac{\big|\mathit{TKE}_{\mathrm{syn}} - \mathit{TKE}_{\mathrm{pre}}\big|}
     {\mathit{TKE}_{\mathrm{pre}}},
\label{eq:TKE_syn}
\end{equation}
where the subscripts `syn' and `pre' denote the synthetic and prescribed values, respectively. 
They are defined as
\begin{equation}
\mathit{TKE}_{\mathrm{syn}} = \tfrac{1}{2}\Big(\overline{u_{\mathrm{syn}}'^{2}} + \overline{v_{\mathrm{syn}}'^{2}} + \overline{w_{\mathrm{syn}}'^{2}}\Big),
\end{equation}
where the overbar indicates spatial averaging over all grid points in the computational domain, and
\begin{equation}
\mathit{TKE}_{\mathrm{pre}} = \tfrac{3}{2}\,u_{\mathrm{pre}}'^{2},
\label{eq:TKE_pre}
\end{equation}
with $u_{\mathrm{pre}}'$ being the prescribed root-mean-square velocity fluctuation under the isotropic turbulence assumption.

Second, at the spectral level, as defined by Saad et al.~\cite{Saad2017}, the relative spectral error is given by:
\begin{equation}
\mathit{Err}_{E(\kappa_{j})} = 
\frac{\big|E_{\mathrm{syn}}(\kappa_{j}) - E_{\mathrm{pre}}(\kappa_{j})\big|}
     {E_{\mathrm{pre}}(\kappa_{j})},
\qquad
\kappa_{\min} \leq \kappa_{j} \leq \kappa_{\max}.
\end{equation}
Here, $E_{\mathrm{pre}}(\kappa_{j})$ is obtained by substituting $\kappa_{j}$ into the target spectrum defined in Eq.~\eqref{eq:spectrum}, while $E_{\mathrm{syn}}(\kappa_{j})$ is computed from the discrete Fourier transform of the synthetic velocity field as
\begin{equation}
E_{\mathrm{syn}}(\kappa_j) = 
\frac{1}{2}\sum_{\bm{\kappa}\in\Omega_j}
\Big(|\hat{u}_{\mathrm{syn}}(\bm{\kappa})|^2 
   + |\hat{v}_{\mathrm{syn}}(\bm{\kappa})|^2 
   + |\hat{w}_{\mathrm{syn}}(\bm{\kappa})|^2\Big),
\end{equation}
where $\hat{u}_{\mathrm{syn}}, \hat{v}_{\mathrm{syn}}, \hat{w}_{\mathrm{syn}}$ are the Fourier coefficients of the synthetic velocity fluctuations and $\Omega_j$ denotes the spectral shell associated with $\kappa_j$.
To assess the overall spectral error, its mean and root-mean-square (rms) values are defined as
\begin{equation}
\mathit{Err}_{E,\mathrm{mean}} = 
\left\langle \mathit{Err}_{E(\kappa_{j})} \right\rangle,
\qquad
\mathit{Err}_{E,\mathrm{rms}} = 
\sqrt{\left\langle \mathit{Err}^{2}_{E(\kappa_{j})} \right\rangle}.
\label{eq:Err_Ek}
\end{equation}
Here, $\mathit{Err}_{E,\mathrm{mean}}$ represents the overall bias trend of the spectral deviation, whereas $\mathit{Err}_{E,\mathrm{rms}}$ reflects the magnitude of the deviations. Together, these three indicators provide a systematic assessment of the consistency between the synthetic turbulence field and the prescribed statistical quantities in terms of both $\mathit{TKE}$ and $E{(\kappa)}$.

\section{Results and Discussion}
\label{sec:3-Results&Discussion}

In a prescribed turbulence energy spectrum, the statistical properties of synthetic turbulence are strongly effected by both the spectral coefficients and the wavenumber range employed for synthesis. In numerical implementation, however, this range is constrained by the grid resolution and domain size, as defined in Eq.~\eqref{eq:KappaRange}. To systematically investigate the errors, this study first investigates the effects of spectral coefficients given in Eq.~\eqref{eq:coef} under the given computational conditions. On this basis, four numerical experiments are conducted:  
(i) modifying the practical wavenumber bounds without changing the grid, to examine the effect of bounds exceeding the theoretical limits on $\mathit{TKE}$ and the spectrum;  
(ii) fixing the domain size while refining the grid, thereby changing the theoretical upper bound;  
(iii) scaling both grid resolution and domain size simultaneously, thereby changing the theoretical lower bound;  
(iv) varying the domain size such that both theoretical bounds change.  
 
\subsection{Effects of Spectral Coefficient Calibration}
\label{sec:3.1}

Before investigating the effects of grid constraints and wavenumber range, it is necessary to first verify the impact of spectral coefficients on the consistency of $\mathit{TKE}$. The baseline case is defined by Eq.~\eqref{eq:spectrum}, with parameters set as $u' = 0.25~\mathrm{m/s}$ and $\varepsilon = 0.5402~\mathrm{m^2/s^3}$, consistent with the settings of Saad et al.~\cite{Saad2017}. With these values, the prescribed turbulent kinetic energy is $\mathit{TKE}_{\mathrm{pre}} = 0.09375~\mathrm{m^2/s^2}$, as outlined in Eq.~\eqref{eq:TKE_pre}.
As described in Section~\ref{sec:2.2}, Saad et al.~\cite{Saad2017} set the von K'arm'an--Pao spectrum coefficients to $\alpha = 1.453$ and $\kappa_{e} = 0.7468/L = 25.82~\mathrm{m^{-1}}$, referred to as case I.
In the present study, the method of Guglielmi et al.~\cite{Guglielmi2025} is also employed, referred to as case II.
Substituting the baseline parameters into Eq.~\eqref{eq:ReL} - Eq.~\eqref{eq:z}, the results are $Re_{L} = 1627$ and $z = 0.000044$. Further substitution into Eq.~\eqref{eq:coef} yields the spectral coefficients $\alpha = 1.561$ and $\kappa_{e} = 22.72~\mathrm{m}^{-1}$.
In summary, the baseline parameters and the two sets of coefficients are listed in Table~\ref{tab:baseline}.

\begin{table}[h!]
\centering
\caption{Baseline parameters and sets of coefficients.}
\label{tab:baseline}
\begin{tabular}{ll}
\hline\hline
\textbf{Parameter category} & \textbf{Value} \\
\hline
Common parameters & $u' = 0.25~\mathrm{m/s},\ 
\varepsilon = 0.5402~\mathrm{m^2/s^3},\ 
\mathit{TKE}_{\mathrm{pre}} = 0.09375~\mathrm{m^2/s^2}$ \\
Case I (First set of coefficients)  & $\alpha = 1.453,\ \kappa_{e} = 25.82~\mathrm{m^{-1}}$, Based on Ref.~\cite{Saad2017}\\
Case II (Second set of coefficients) & $\alpha = 1.561,\ \kappa_{e} = 22.72~\mathrm{m^{-1}}$, From Eq.~\eqref{eq:coef}\\
\hline\hline
\end{tabular}
\end{table}

Using these two sets of coefficients in Eq.~\eqref{eq:spectrum}, the $\mathit{TKE}$ derived from the turbulence spectrum is computed using the trapezoidal rule:
\begin{equation}
\mathit{TKE} \;\approx\; 
\sum_{i=1}^{N-1} \frac{E(\kappa_i) + E(\kappa_{i+1})}{2}\,\Delta \kappa_i,
\label{eq:TKE_integ}
\end{equation}
where $N=10^{2},\,10^{3},\,10^{4},\,10^{5}$ equally spaced quadrature points are tested. The lower and upper bounds are set to $0.01\kappa_{e}$ and $10\kappa_{\eta}$, respectively. The resulting $\mathit{TKE}$ and relative errors are summarized in Table~\ref{tab:tke_integration}. When $N$ reaches $10^{4}$, the integral essentially converges. For case I, the computed $\mathit{TKE}$ is $0.08667$ with a relative error of $7.551\%$; for case II, the computed $\mathit{TKE}$ is $0.09375$ with an error below $0.0001\%$. These results indicate that the calibrated coefficients enforce consistency of $\mathit{TKE}$ with the theoretical value, whereas the remaining discrepancies are mainly attributable to the finite wavenumber range imposed by numerical discretization.

\begin{table}[htbp]
  \centering
  \caption{Effect of numerical integration points on $\mathit{TKE}$ calculated from $E(\kappa)$}
  \label{tab:tke_integration}
  \begin{tabular}{cccccc}
    \hline\hline  
    \multirow{2}{*}{$\mathit{TKE}_{\mathrm{pre}}$ ($\mathrm{m^2/s^2}$)} & \multirow{2}{*}{Integration Points, $N$} & \multicolumn{2}{c}{First Set of Coefficients} & \multicolumn{2}{c}{Second Set of Coefficients} \\
    \cline{3-4} \cline{5-6} 
    & & $\mathit{TKE}_{\mathrm{syn}}$($\mathrm{m^2/s^2}$) & $\mathit{Err_{TKE}}$ & $\mathit{TKE}_{\mathrm{syn}}$($\mathrm{m^2/s^2}$) & $\mathit{Err_{TKE}}$ \\
    \hline  
    \multirow{4}{*}{0.09375} & $10^2$ & 0.02001 & 78.61\% & 0.01982 & 78.86\% \\
    & $10^3$ & 0.08456 & 9.799\% & 0.08892 & 5.149\% \\
    & $10^4$ & 0.08667 & 7.551\% & 0.09375 & 0.000067\% \\
    & $10^5$ & 0.08667 & 7.551\% & 0.09375 & 0.000001\% \\
    \hline\hline 
  \end{tabular}
\end{table}

In the preceding comparison, the two sets of coefficients are examined by directly integrating the prescribed turbulence spectrum using Eq.~\eqref{eq:spectrum} and Eq.~\eqref{eq:TKE_integ} to check whether the resulting energy equals the target $\mathit{TKE}$. 
Here, the same two sets are further applied to synthetic turbulence generation, and the statistical results are evaluated according to Eq.~\eqref{eq:TKE_syn} – Eq.~\eqref{eq:Err_Ek}. 
The grid is configured with $128^{3}$ points, and the computational domain has a length of $0.18\pi~\mathrm{m}$, with a linear distribution of $10^{3}$ modes, consistent with Saad et al.~\cite{Saad2017}. This linear modal distribution is also consistent with the results from the numerical integration tests mentioned above. The random seed is fixed at 0. 
The statistical results are summarized in Table~\ref{tab:tke_syn}. 
The $\mathit{Err}_{\mathit{TKE}}$ difference between case I and case II is approximately $7.14\%$, which aligns with the converged $\mathit{Err}_{\mathit{TKE}}$ value of $7.551\%$ from Table~\ref{tab:tke_integration}.
However, the synthetic turbulence generated with case II exhibits a $\mathit{Err}_{\mathit{TKE}}$ of $10.14\%$, while the mean and RMS spectral errors remain nearly unchanged between the two cases.
These statistical results indicate that coefficient calibration has a primary but limited effect on $\mathit{TKE}$ consistency, with negligible impact on spectral consistency. Therefore, the second set of coefficients of case II is adopted in the following sections to facilitate a systematic investigation of the effects of wavenumber bounds in synthetic turbulence.

\begin{table}[h!]
\centering
\caption{Comparison of synthetic turbulence statistics using the two sets of coefficients.}
\label{tab:tke_syn}
\begin{tabular}{lcccc}
\hline\hline
Case Name & $\mathit{TKE}_{\mathrm{syn}}$ & $\mathit{Err}_{\mathit{TKE}}$ & $\mathit{Err}_{E,\mathrm{mean}}$ & $\mathit{Err}_{E,\mathrm{rms}}$ \\
\hline
Case I (First set of coefficients)  & 0.07755 & 17.28\% & 6.143\% & 6.717\% \\
Case II (Second set of coefficients) & 0.08424 & 10.14\% & 6.606\% & 7.200\% \\
\hline\hline
\end{tabular}
\end{table}

\subsection{Effects of Extending Wavenumber Bounds Beyond Theoretical Limits}
\label{sec:3.2}

As mentioned above, Table~\ref{tab:wavenumber} summarizes different approaches in the literature for choosing the wavenumber range. For homogeneous isotropic turbulence synthesized on a Cartesian grid, the wavenumber range defined by Eq.~\eqref{eq:KappaRange} is the one most consistent with sampling theory. However, in practical applications, many studies also define the wavenumber range based on physical scales~\cite{Bechara1994,Bailly1999}. To examine the impact of this difference on the statistical consistency of synthetic turbulence, this subsection considers the same grid while directly changing the actual lower and upper wavenumber bounds used in turbulence synthesis, and analyzes their influence on $\mathit{TKE}$ and $E{(\kappa)}$.

Based on the baseline case in Section~\ref{sec:3.1} and Table~\ref{tab:baseline}, the grid resolution is fixed at $128^{3}$ points with a computational domain length of $0.18\pi~\mathrm{m}$. Both $\kappa_{\min}=11.11~\mathrm{m}^{-1}$ and $\kappa_{\max}=711.1~\mathrm{m}^{-1}$ can be determined from Eq.~\eqref{eq:KappaRange}.
The wavenumber range $[\kappa_{\min}, \kappa_{\max}]$ is then modified in turn to $[0.1\kappa_{\min}, \kappa_{\max}]$, $[\kappa_{\min}, 10\kappa_{\max}]$, and $[0.1\kappa_{\min}, 10\kappa_{\max}]$. In addition, the physically relevant wavenumbers are $\kappa_{e}=22.72~\mathrm{m}^{-1}$ and $\kappa_{\eta}=4821~\mathrm{m}^{-1}$, respectively.

\begin{figure}[htbp]
  \centering
  \includegraphics[width=0.6\linewidth]{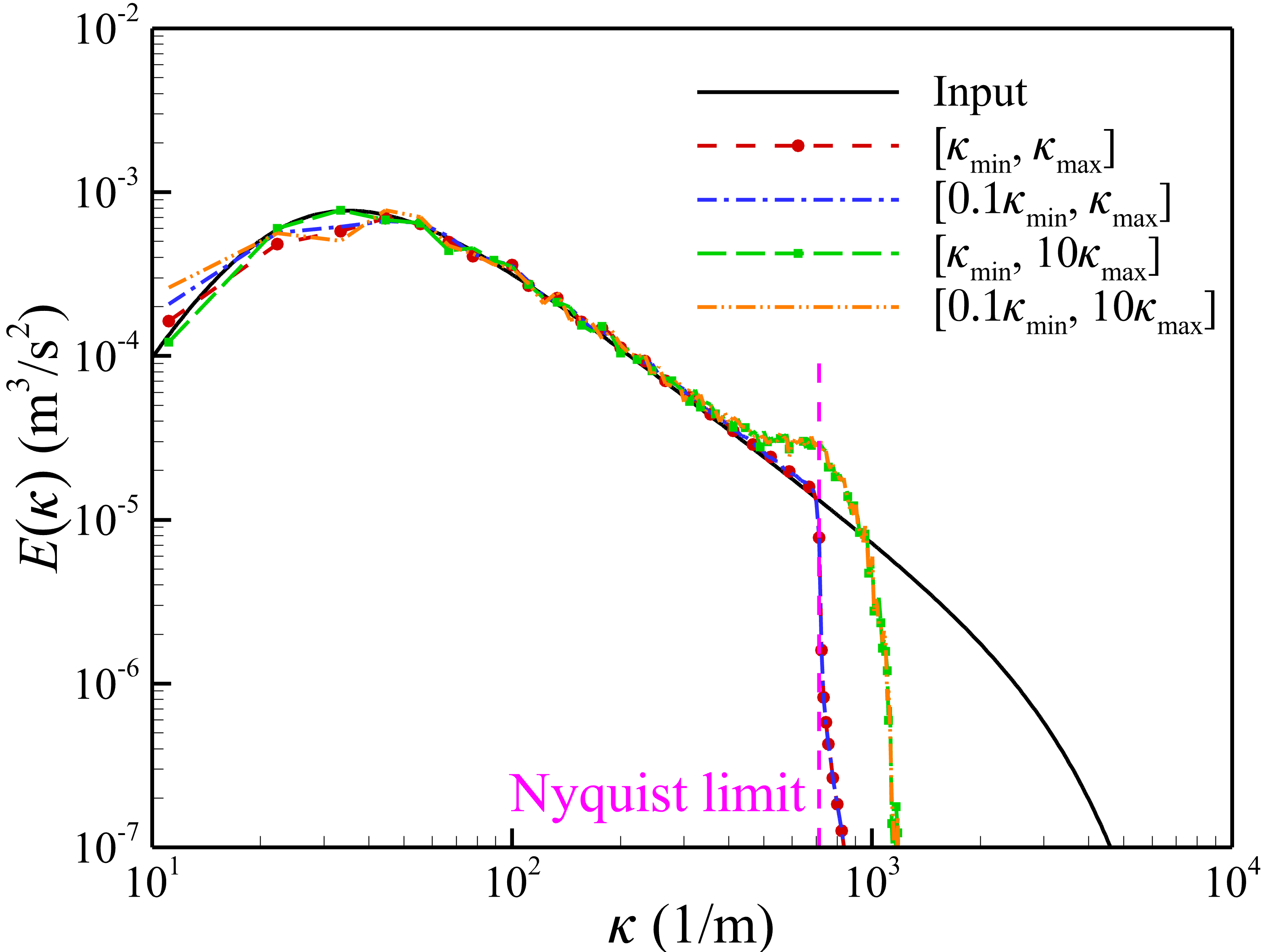}
  \caption{Comparison of turbulence energy spectra under different actual wavenumber bounds.}
  \label{fig:kbound}
\end{figure}

The statistical results for the synthetic turbulence under the four wavenumber ranges are given in Table~\ref{tab:kbound}, and the corresponding energy spectra are shown in Fig.~\ref{fig:kbound}. When the upper bound is extended to $10\kappa_{\max}$, $\mathit{Err}_{\mathit{TKE}}$ decreases significantly to 1.239\%, almost eliminating the overall energy bias; but the high-wavenumber part of the spectrum rises markedly, leading to a sharp increase in both the mean and RMS spectral errors to 28.50\% and 41.15\%, respectively. 
These statistical results indicate that the upper bound governs the representation of total energy, and that increasing this bound helps to reduce the $\mathit{TKE}$ error.
However, if the actual upper bound exceeds the Nyquist limit, high-wavenumber energy spuriously accumulates within the resolvable range due to aliasing, thereby distorting the low-wavenumber spectrum. 
When reviewing Table~\ref{tab:wavenumber}, if the approach of Béchara et al.~\cite{Bechara1994} is adopted, in which grid constraints are neglected and the bounds are defined solely from physical scales, it may reduce the $\mathit{TKE}$ error but introduces additional spectral inconsistencies. Therefore, the subsequent analysis is restricted to the theoretical wavenumber bounds determined by the grid constraints.

\begin{table}[htbp]
  \centering
  \caption{Statistical results and error comparison of synthetic turbulence under different actual wavenumber bounds.}
  \label{tab:kbound}
  \begin{tabular}{ccccc}
    \hline\hline
    Wavenumber range & $\mathit{TKE}_{\mathrm{syn}}$($\mathrm{m^2/s^2}$) & $\mathit{Err}_{\mathit{TKE}}$ & $\mathit{Err}_{E,\mathrm{mean}}$ & $\mathit{Err}_{E,\mathrm{rms}}$ \\
    \hline
    $[\kappa_{\min}, \kappa_{\max}]$ & 0.08424 & 10.14\% & 6.606\% & 7.200\% \\
    $[0.1\kappa_{\min}, \kappa_{\max}]$ & 0.08671 & 7.509\% & 6.509\% & 7.093\% \\
    $[\kappa_{\min}, 10\kappa_{\max}]$ & 0.09491 & 1.239\% & 27.55\% & 40.34\% \\
    $[0.1\kappa_{\min}, 10\kappa_{\max}]$ & 0.09447 & 0.7630\% & 28.50\% & 41.15\% \\
    \hline\hline
  \end{tabular}
  
  \vspace{0.3em}
  \noindent Note: $\kappa_{\min}=11.11~\mathrm{m^{-1}}$ and $\kappa_{\max}=711.1~\mathrm{m^{-1}}$, both determined from Eq.~\eqref{eq:KappaRange}.
\end{table}

\subsection{Effects of Varying the Theoretical Upper Wavenumber Bound}
\label{sec:3.3}

\begin{figure}[!htb]
    \centering
    \includegraphics[width=0.6\linewidth]{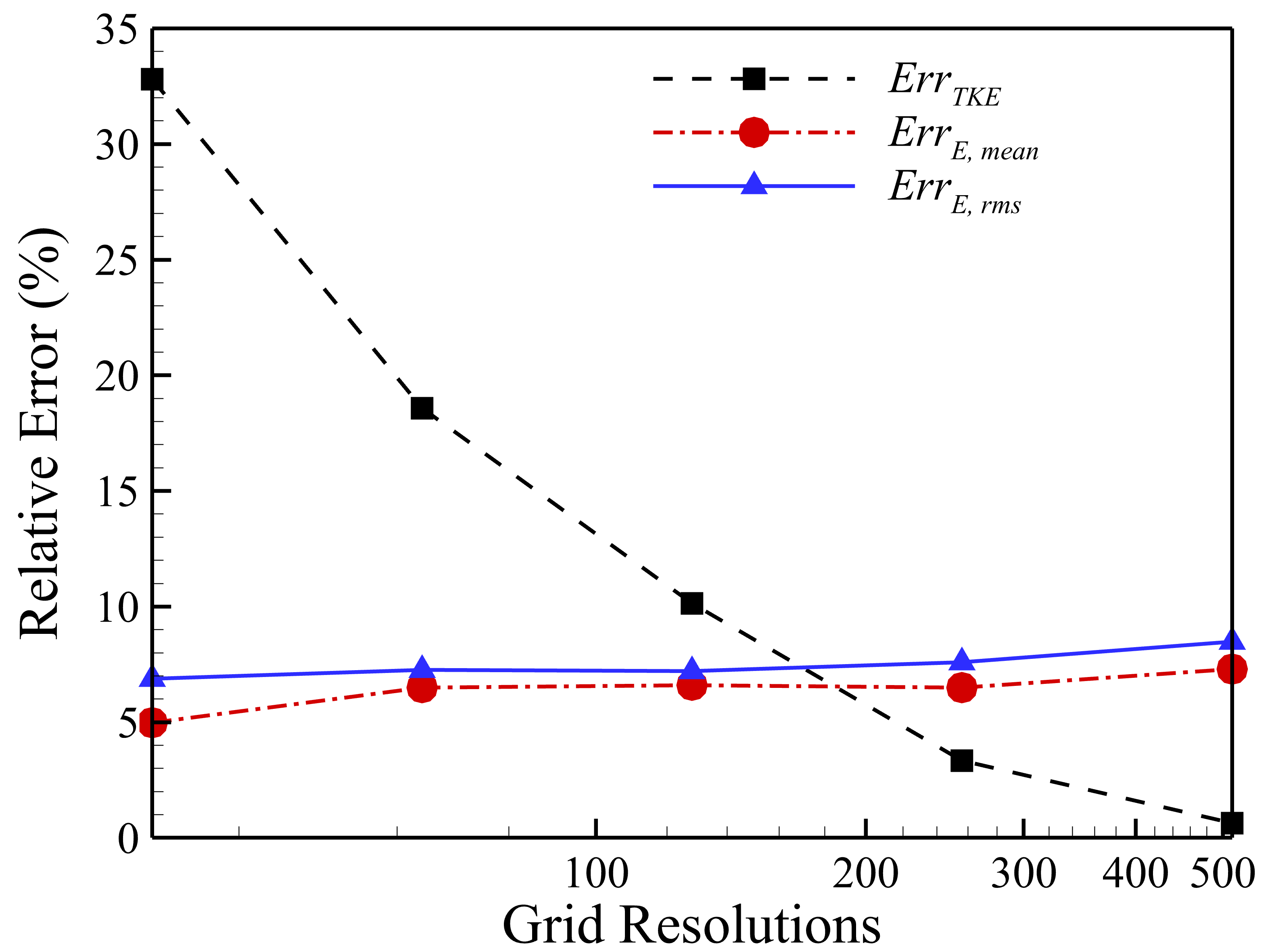} 
    \caption{Relative errors with varying theoretical upper wavenumber bound.}
    \label{fig:error_kmax}
\end{figure}

Keeping the computational domain size fixed, the grid resolution is gradually increased, which is equivalent to increasing the theoretical upper wavenumber bound $\kappa_{\max}$. The statistical results of synthetic turbulence for five different upper bounds are given in Table~\ref{tab:kmax}, and the variations of the three errors with grid resolution are shown in Fig.~\ref{fig:error_kmax}. It can be seen that $\mathit{Err}_{\mathit{TKE}}$ decreases markedly with increasing grid resolution and theoretical $\kappa_{\max}$. When the grid is refined from $32^{3}$ to $512^{3}$, $\mathit{Err}_{\mathit{TKE}}$ decreases from 32.8\% to 0.64\%, exhibiting a clear convergence trend. This indicates that the consistency of $\mathit{TKE}$ depends strongly on the high-wavenumber range. 

\begin{figure}[!htb]
    \centering
    \includegraphics[width=0.6\linewidth]{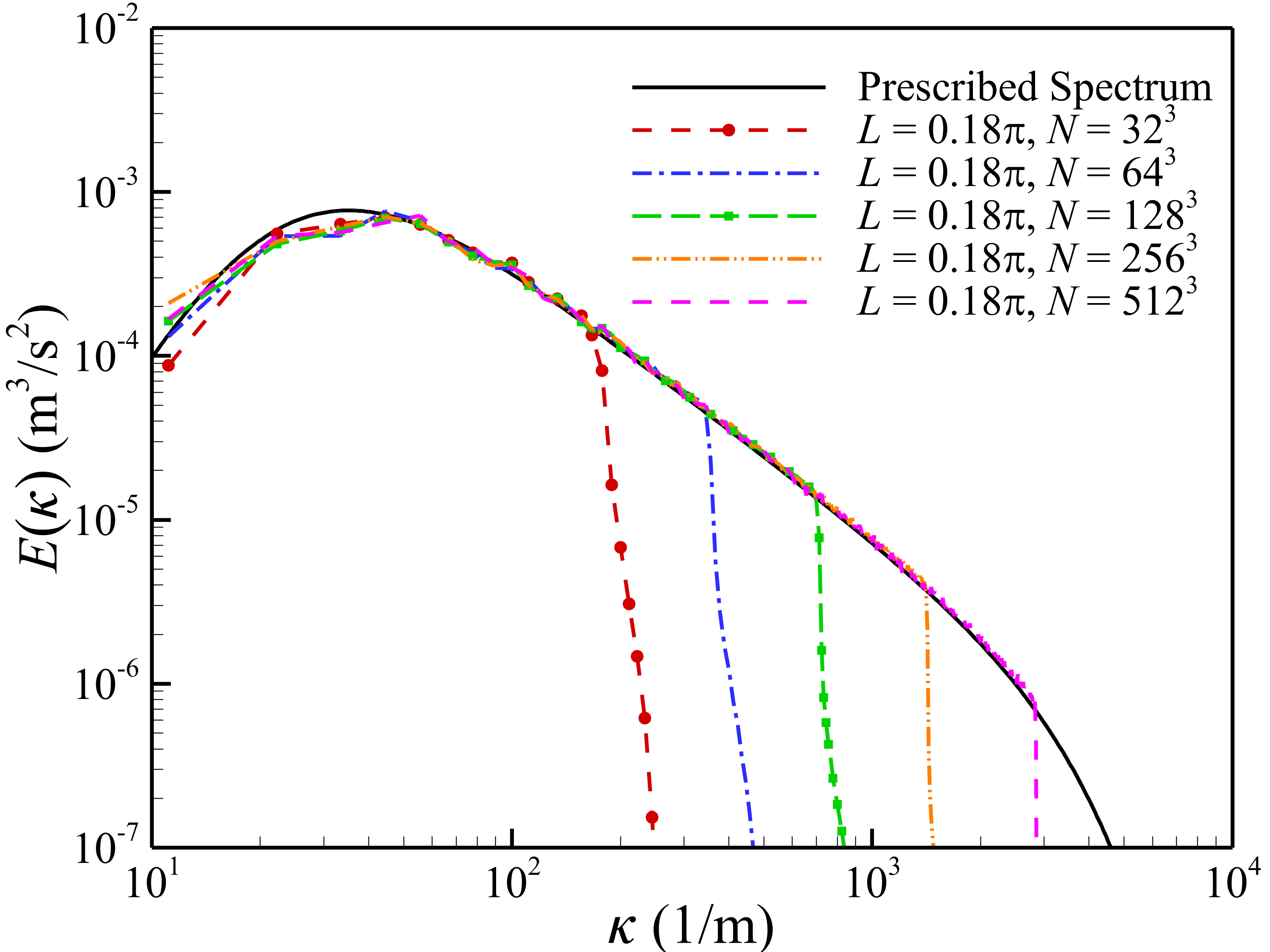} 
    \caption{Energy spectrum distributions with varying theoretical upper wavenumber bound.}
    \label{fig:spectrum_kmax}
\end{figure}

Figure~\ref{fig:spectrum_kmax} further illustrates the spectral distributions of synthetic turbulence at different grid resolutions and upper bounds. The synthetic spectra closely follow the prescribed distribution over a larger range in the mid-to-high wavenumber range as the upper wavenumber bound increases. Under the present conditions, the Kolmogorov wavenumber is $\kappa_{\eta} \simeq 4821~\mathrm{m^{-1}}$. For the $512^{3}$ grid, $\kappa_{\max}/\kappa_{\eta} = 0.59$, indicating partial representation of the high-wavenumber range, which leads to a reduction of $\mathit{Err}_{\mathit{TKE}}$ to as low as 0.64\%.

On the other hand, Table~\ref{tab:kmax} also shows that $\mathit{Err}_{E,\mathrm{mean}}$ and $\mathit{Err}_{E,\mathrm{rms}}$ remain within 4.981-8.486\% and exhibit little improvement with increasing $\kappa_{\max}$, consistent with the results of Saad et al.~\cite{Saad2017}. Figure~\ref{fig:spectrum_kmax} also indicates that spectral deviations are concentrated around the peak region and are only weakly affected by the upper wavenumber bound. This confirms that the upper bound primarily governs $\mathit{TKE}$ consistency, whereas the low-wavenumber spectral discrepancies are largely unaffected.

\begin{table}[htbp]
  \centering
  \caption{Statistical results and error comparison of synthetic turbulence with varying theoretical upper wavenumber bound.}
  \label{tab:kmax}
  \begin{tabular}{ccccccc}
    \hline\hline
    Domain size, $L$ (m) & Grid resolution & $[\kappa_{\min}, \kappa_{\max}]$ & $\mathit{TKE}_{\mathrm{syn}}$ ($\mathrm{m^2/s^2}$) & $\mathit{Err}_{\mathit{TKE}}$ & $\mathit{Err}_{E,\mathrm{mean}}$ & $\mathit{Err}_{E,\mathrm{rms}}$ \\
    \hline
    $0.18\pi$ & $32^3$  & [11.11, 177.9]  & 0.06299 & 32.81\% & 4.981\% & 6.886\% \\
    $0.18\pi$ & $64^3$  & [11.11, 355.9]  & 0.07634 & 18.57\% & 6.495\% & 7.258\% \\
    $0.18\pi$ & $128^3$ & [11.11, 711.1]  & 0.08424 & 10.14\% & 6.606\% & 7.200\% \\
    $0.18\pi$ & $256^3$ & [11.11, 1422]   & 0.09062 & 3.34\%  & 6.495\% & 7.586\% \\
    $0.18\pi$ & $512^3$ & [11.11, 2844]   & 0.09315 & 0.64\%  & 7.293\% & 8.486\% \\
    \hline\hline
  \end{tabular}
\end{table}

\subsection{Effects of Varying the Theoretical Lower Wavenumber Bound}
\label{sec:3.4}

\begin{figure}[!htb]
    \centering
    \includegraphics[width=0.6\linewidth]{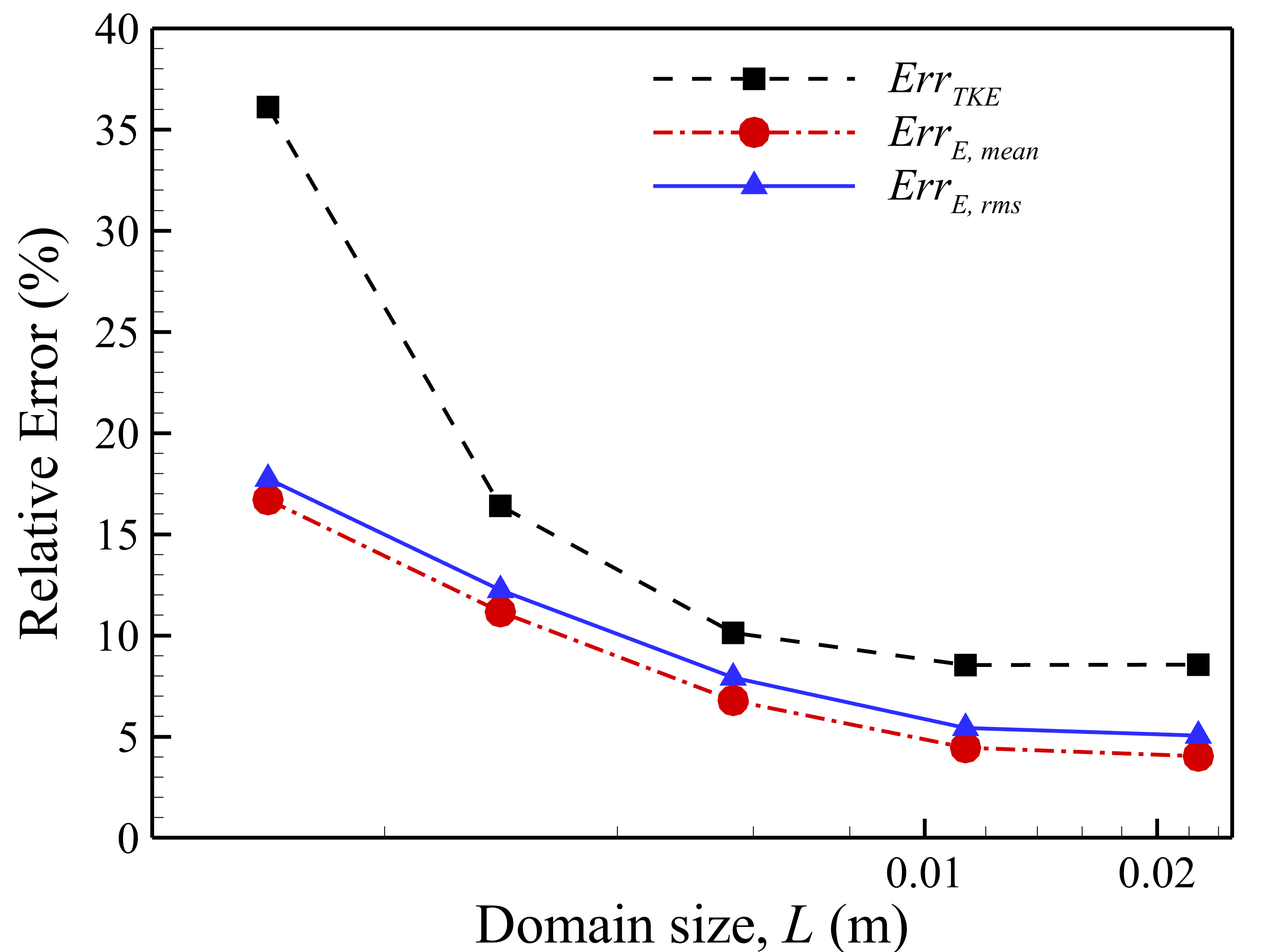} 
    \caption{Relative errors with varying theoretical lower wavenumber bound.}
    \label{fig:error_kmin}
\end{figure}

To further examine the influence of the lower bound of the wavenumber range in synthetic turbulence, the grid spacing is kept constant while the computational domain size and grid resolution are scaled proportionally. In this way, $\kappa_{\max}$ remains unchanged, and only the theoretical lower bound $\kappa_{\min}$ is modified. The statistical results corresponding to five different lower bounds are summarized in Table~\ref{tab:kmin}, and the variation of the three errors with $\kappa_{\min}$ is presented in Fig.~\ref{fig:error_kmin}. It is shown that as $\kappa_{\min}$ decreases, $\mathit{Err}_{\mathit{TKE}}$ decreases significantly from 36.12\% to about 8\%. At the same time, both $\mathit{Err}_{E,\mathrm{mean}}$ and $\mathit{Err}_{E,\mathrm{rms}}$ decrease from approximately 17\% to below 5\%. Compared with the results of Saad et al.~\cite{Saad2017}, enlarging the computational domain and thereby lowering $\kappa_{\min}$ reduces the spectral error below 5\% and shows a stable convergence trend.

\begin{figure}[!htb]
    \centering
    \includegraphics[width=0.6\linewidth]{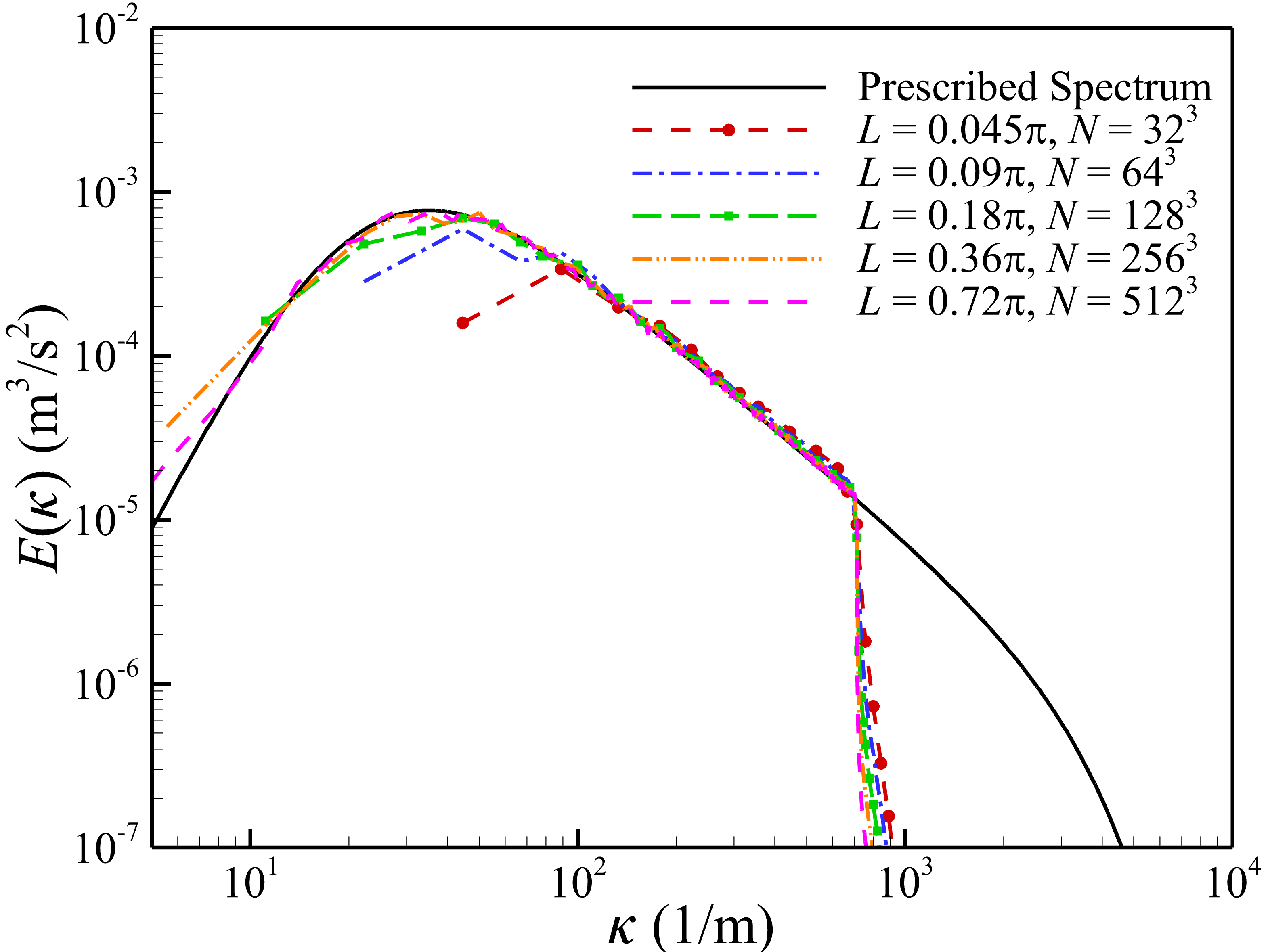} 
    \caption{Energy spectrum distributions with varying theoretical lower wavenumber bound.}
    \label{fig:spectrum_kmin}
\end{figure}

Fig.~\ref{fig:spectrum_kmin} compares the energy spectrum distributions corresponding to different theoretical lower bounds. The results show that reducing $\kappa_{\min}$ primarily improves the consistency of the spectrum in the low-wavenumber range. For the present case with $\kappa_{e} = 22.72~\mathrm{m^{-1}}$, the ratio $\kappa_{\min}/\kappa_{e}$ decreases from 1.96 to 0.12. When $\kappa_{\min}$ is relatively large, the low-wavenumber energy in the energy-containing range is truncated, leading to substantial deviation in the overall spectral distribution. When $\kappa_{\min}$ is sufficiently small, spectral consistency is improved because the $k^{4}$ rising region of the spectrum (Eq.~\eqref{eq:spectrum}) is fully represented in the synthetic field, which reduces systematic deviations associated with truncated low-wavenumber contributions. These results clearly show that the computational domain size and the low-wavenumber coverage of synthetic turbulence play a decisive role in the consistency of $E(\kappa)$.

\begin{table}[htbp]
  \centering
  \caption{Statistical results and error comparison of synthetic turbulence with varying theoretical lower wavenumber bound.}
  \label{tab:kmin}
  \begin{tabular}{ccccccc}
    \hline\hline
    Domain size, $L$ (m) & Grid resolution & $[\kappa_{\min}, \kappa_{\max}]$ & $\mathit{TKE}_{\mathrm{syn}}$ ($\mathrm{m^2/s^2}$) & $\mathit{Err}_{\mathit{TKE}}$ & $\mathit{Err}_{E,\mathrm{mean}}$ & $\mathit{Err}_{E,\mathrm{rms}}$ \\
    \hline
    0.045$\pi$ & $32^3$   & [44.44, 711.1] & 0.05989 & 36.12\% & 16.71\% & 17.75\% \\
    0.09$\pi$  & $64^3$   & [22.22, 711.1] & 0.07837 & 16.40\% & 11.16\% & 12.25\% \\
    0.18$\pi$  & $128^3$  & [11.11, 711.1] & 0.08424 & 10.14\% & 6.778\% & 7.920\% \\
    0.36$\pi$  & $256^3$  & [5.556, 711.1] & 0.08575 & 8.537\% & 4.448\% & 5.430\% \\
    0.72$\pi$  & $512^3$  & [2.778, 711.1] & 0.08572 & 8.567\% & 4.036\% & 5.054\% \\
    \hline\hline
  \end{tabular}
\end{table}

\subsection{Effects of Varying Both Theoretical Wavenumber Bounds}
\label{sec:3.5}

\begin{figure}[!htb]
    \centering
    \includegraphics[width=0.6\linewidth]{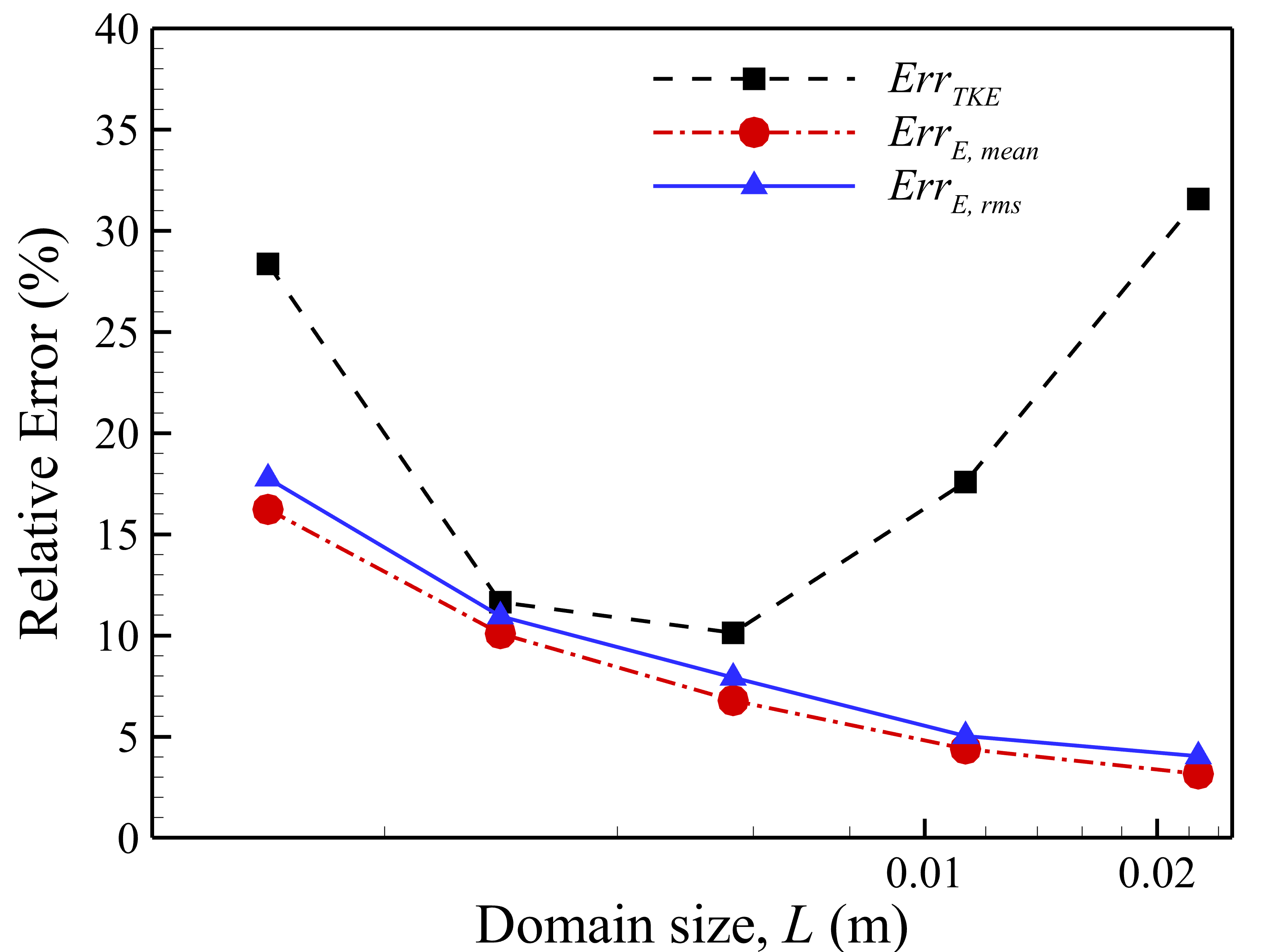} 
    \caption{Relative errors with varying both theoretical lower and upper wavenumber bounds.}
    \label{fig:error_kboth}
\end{figure}

Different from the previous subsection, here the grid resolution is fixed at $N=128^{3}$, while the computational domain size $L$ is varied. This simultaneously alters both $\kappa_{\max}$ and $\kappa_{\min}$, thereby enabling investigation of the combined effects of changing the upper and lower wavenumber bounds. The statistical results for five different wavenumber ranges are listed in Table~\ref{tab:kjoint}, and the variation of the three errors is shown in Fig.~\ref{fig:error_kboth}. It can be observed that $\mathit{Err}_{\mathit{TKE}}$ first decreases and then increases with increasing domain size. When $L$ is small, $\kappa_{\min}$ is large, truncating part of the energy-containing range and resulting in underestimated $\mathit{TKE}$. As $L$ increases, the energy-containing scales are gradually captured, and $\mathit{Err}_{\mathit{TKE}}$ decreases to about 10\%. For sufficiently large $L$, however, $\kappa_{\max}$ also increases substantially, and the high-wavenumber range becomes insufficiently represented in the synthetic field, again leading to underestimation of $\mathit{TKE}$.

\begin{figure}[!htb]
    \centering
    \includegraphics[width=0.6\linewidth]{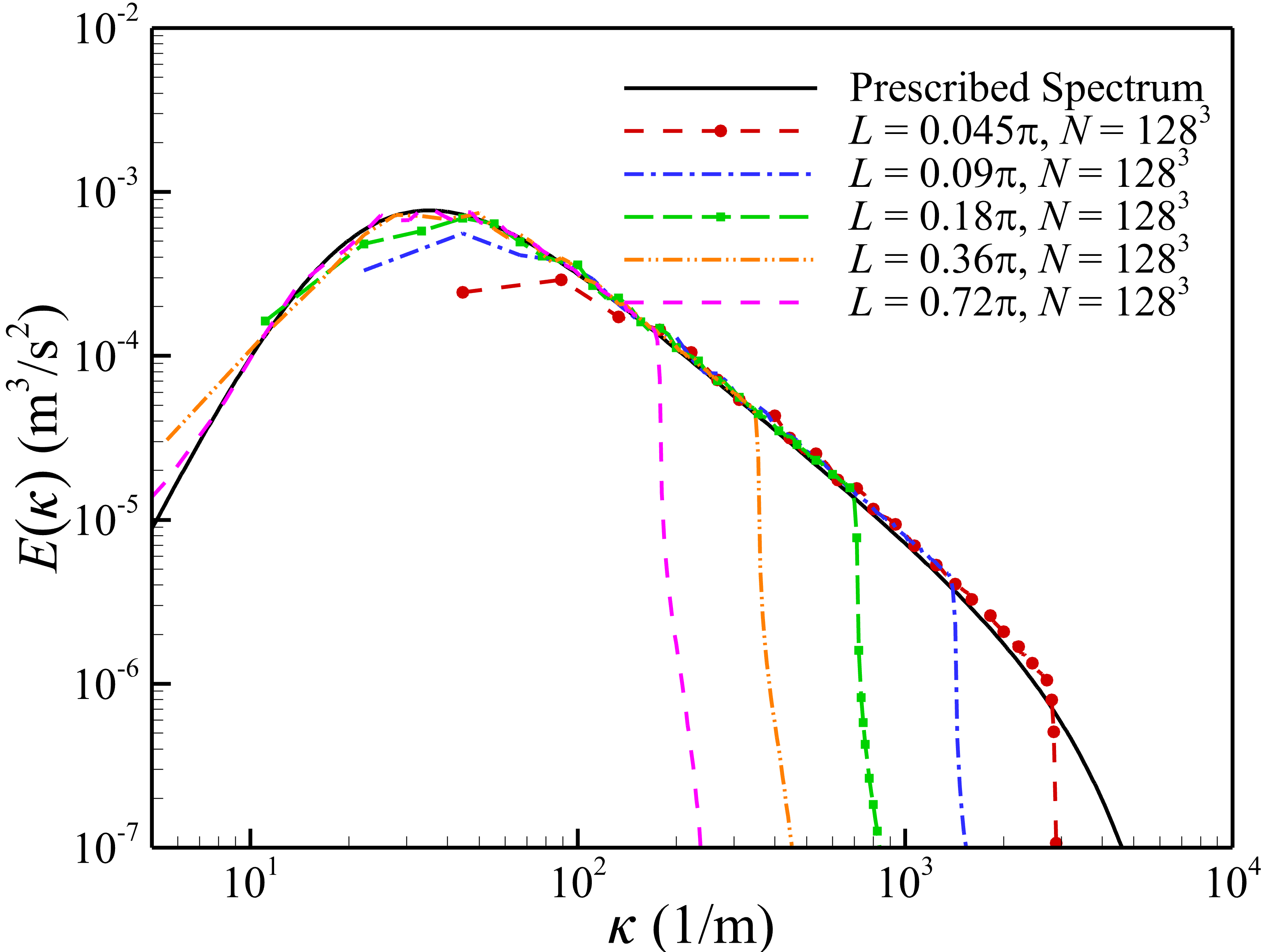} 
    \caption{Energy spectrum distributions with varying both theoretical lower and upper wavenumber bounds.}
    \label{fig:spectrum_kboth}
\end{figure}

In contrast, the trend of spectral errors do not fully follow that of $\mathit{TKE}$: both $\mathit{Err}_{E,\mathrm{mean}}$ and $\mathit{Err}_{E,\mathrm{rms}}$ decrease nearly linearly with increasing domain size. Figure~\ref{fig:spectrum_kboth} shows the corresponding energy spectra. The results reveal that spectral error mainly originates from insufficient coverage of the low-wavenumber region, and enlarging the domain size effectively alleviates this issue, even when the high-wavenumber region is not fully captured. This also directly addresses the observation of Saad et al.~\cite{Saad2017} regarding the persistent energy deficit near the integral scale, which cannot be mitigated by increasing the number of modes or adopting logarithmic distributions. The present study shows that alleviating this deficit necessitates expansion of the lower bound of the wavenumber range.

Consistent with Section~\ref{sec:3.4}, these findings indicate that enlarging the computational domain enhances low-wavenumber coverage and improves $E(\kappa)$ consistency. 
However, if the grid resolution is not increased accordingly, high-wavenumber energy will inevitably be underestimated, which in turn affects the consistency of $\mathit{TKE}$. Therefore, when the focus is on reproducing $E{(\kappa)}$, a larger domain size is preferable, while ensuring consistency of $\mathit{TKE}$ requires balancing between domain size and resolution.

\begin{table}[htbp]
  \centering
  \caption{Statistical results and error comparison of synthetic turbulence with varying both theoretical wavenumber bounds.}
  \label{tab:kjoint}
  \begin{tabular}{ccccccc}
    \hline\hline
    Domain size, $L$ (m) & Grid resolution & $[\kappa_{\min}, \kappa_{\max}]$ & $\mathit{TKE}_{\mathrm{syn}}$ ($\mathrm{m^2/s^2}$) & $\mathit{Err}_{\mathit{TKE}}$ & $\mathit{Err}_{E,\mathrm{mean}}$ & $\mathit{Err}_{E,\mathrm{rms}}$ \\
    \hline
    0.045$\pi$ & $128^3$ & [44.44, 2844] & 0.06715 & 28.37\% & 16.23\% & 17.77\% \\
    0.09$\pi$  & $128^3$ & [22.22, 1422] & 0.08283 & 11.65\% & 10.09\% & 10.95\% \\
    0.18$\pi$  & $128^3$ & [11.11, 711.1] & 0.08424 & 10.14\% & 6.778\% & 7.920\% \\
    0.36$\pi$  & $128^3$ & [5.556, 355.6] & 0.07726 & 17.59\% & 4.401\% & 5.031\% \\
    0.72$\pi$  & $128^3$ & [1.389, 177.8] & 0.06415 & 31.57\% & 3.154\% & 4.032\% \\
    \hline\hline
  \end{tabular}
\end{table}

\section{Conclusions}
\label{sec:4-Conclusion}
This study systematically evaluates the statistical consistency between synthetic turbulence and prescribed targets under the constraints of finite grid and bounded wavenumber ranges when using the Random Fourier Method (RFM). In order to clarify the origin of error, the investigation evaluates discrepancies in the turbulent kinetic energy ($\mathit{TKE}$) and the turbulence energy spectrum($E(\kappa)$). The effect of spectral coefficient calibration on $\mathit{TKE}$ is first clarified. On this basis, a series of numerical experiments are conducted to systematically identify the effects of different lower– and upper–wavenumber combinations on the consistency of $\mathit{TKE}$ and $E(\kappa)$.
The main observations and conclusions are summarized as follows:  
\begin{enumerate}
    \item Calibration of the spectral coefficients is a prerequisite for achieving $\mathit{TKE}$ consistency. With an unrestricted wavenumber range, the recalibrated coefficients eliminate the discrepancy between the computed and prescribed $\mathit{TKE}$.  
    \item The upper wavenumber bound primarily controls the $\mathit{TKE}$ consistency. Increasing grid resolution, thereby raising the theoretical upper bound, significantly reduces the $\mathit{TKE}$ error. 
    \item The lower wavenumber bound governs the fidelity of the low-wavenumber spectrum. Enlarging the computational domain and thereby reducing the lower bound improves $E(\kappa)$ consistency in the low-wavenumber range.
    \item The wavenumber range of synthetic turbulence should be selected with consideration of the grid constrains. When the bounds are prescribed only based on physical scales but in practice exceed the grid-constrained limit, aliasing effects occur, leading to a substantial increase in $E(\kappa)$ error.
\end{enumerate}

This work clarifies the error origin of RFM with wavenumber bounds and provides practical guidance for wavenumber selection, thus enhancing its reliability in computational fluid dynamics and computational aeroacoustics applications. To eliminate the potential influence of mode discretizations, a sufficiently large number of Fourier modes was adopted. On the basis of the present findings, future investigations will focus on the effects of both the number and distribution of modes within fixed wavenumber bounds.

\section*{Acknowledgment}
    This work is supported by NSFC the Excellence Research Group Program for multiscale problems in nonlinear mechanics (Grant No. 12588201), the National Natural Science Foundation of China (Nos. 12425207, 92252203 and 12102439), the Chinese Academy of Sciences Project for Young Scientists in Basic Research (Grant No. YSBR-087), and the Strategic Priority Research Program of Chinese Academy of Sciences (Grant No. XDB0620102). The authors acknowledge the assistance of ChatGPT in improving readability, grammar, and language in this work.

\bibliography{refs}

\end{document}

%% file: annotation.tex
\definecolor{lightblue}{rgb}{.90,.95,1}
\definecolor{darkgreen}{rgb}{0,.5,0.5}
\definecolor{lightgreen}{rgb}{.90,1,0.90}

